\documentclass[prb,showpacs,preprintnumbers,twocolumn,amsmath,amssymb,superscriptaddress]{revtex4-1}

\usepackage[T1]{fontenc}
\usepackage[applemac]{inputenc}
\usepackage[english]{babel}
\usepackage{graphicx}
\usepackage{amssymb}
\usepackage{amsmath}
\usepackage{overpic}
\usepackage{color}

\newcommand{\be}{\begin{equation}}
\newcommand{\ee}{\end{equation}}
\newcommand{\bea}{\begin{eqnarray}}
\newcommand{\eea}{\end{eqnarray}}

\def\a{\alpha}
\def\b{\beta}
\def\e{\varepsilon}
\def\d{\delta}
\def\g{\gamma}

\renewcommand{\o}{\omega}

\def\s{\sigma}

\def\D{\Delta}
\renewcommand{\O}{\Omega}

\def\ra{\rightarrow}

\def\pll{\parallel}

\def\pd{\partial}
\def\nb{\nabla}
\def\bk{{\bf k}}
\def\bj{{\bf j}}

\def\bq{{\bf q}}

\def\bA{{\bf A}}

\def\bE{{\bf E}}

\def\bJ{{\bf J}}

\def\nn{\nonumber}
\def\lb{\label}
\def\pref#1{(\ref{#1})}

\newcount\bozza \bozza=0
\ifnum\bozza=1
\newdimen\shift \shift=-2truecm
\def\lb#1{%
{\label{#1}\rlap{\kern\shift{$\scriptstyle#1$}}}}
\else\def\lb#1{\label{#1}} \fi

\def\insertplot#1#2#3#4{\begin{minipage}{#2}
\vbox {\hbox to #1 {\vbox to #2 {\vfil%
\includegraphics{#4.pst}#3}}}
\end{minipage}}

\def\ins#1#2#3{\vbox to0pt{\kern-#2 \hbox{\kern#1 #3}\vss}\nointerlineskip}

\def\nn{\nonumber}

\begin{document}
\title{Polarization dependence of the third-harmonic generation\\ in multiband superconductors}
\author{T. Cea}
\affiliation{IMDEA Nanoscience, C/Faraday 9, 28049 Madrid, Spain}
\affiliation{Graphene Labs, Fondazione Istituto Italiano di Tecnologia, Via Morego, 16163 Genova, Italy}
\affiliation{ISC-CNR and Dep. of Physics, Sapienza University of Rome, P.le A. Moro 5, 00185 Rome, Italy}
  \author{P. Barone}
\affiliation{SPIN-CNR, via Vetoio, 67100 L'Aquila, Italy}
\author{C. Castellani}
\affiliation{ISC-CNR and Dep. of Physics, Sapienza University of Rome, P.le A. Moro 5, 00185 Rome, Italy}
\author{L. Benfatto}
\affiliation{ISC-CNR and Dep. of Physics, Sapienza University of Rome, P.le A. Moro 5, 00185 Rome, Italy}
  \email{lara.benfatto@roma1.infn.it}
\date{\today}

\begin{abstract}
In a superconductor the third-harmonic generation (THG) of a strong THz pulse is enhanced below $T_c$ by the resonant excitation of lattice-modulated charge fluctuations (LCF), which modulate the response according to the polarization of the field. Here we compute the THG within a multiband model for the prototype NbN superconductor. We show that the non-resonant contribution coming from the instantaneous electronic response and the finite width of the pulse significantly suppress the polarization dependence of the  signal,  challenging its observation in real systems. 
\end{abstract}

\maketitle
\section{Introduction}

The use of intense THz field has recently opened  the avenue to an alternative way to detect and excite low-energy excitations in solids\cite{review1,review2}. In particular, THz spectroscopy at high fields is an excellent tool to address the physics of superconducting (SC) systems, where the relevant single-particle and collective degrees of freedom can be resonantly excited exactly in this frequency range\cite{giannetti_review,cavalleri_review}. While the understanding of pump-probe protocols could involve non-equilibrium processes, transmission experiments can be understood by equilibrium response. On this respect, the observation\cite{shimano14,shimano17}  of an enhanced third-harmonic of the incident field when the pump frequency matches the gap value $\Delta_0$ has triggered the theoretical investigation of non-linear optical effects in superconductors\cite{aoki_prb15,cea_prb16,aoki_sc_prb16,cea_raman_prb16,aoki_mgb2_prb17,biancalana_prb17}.

Despite the initial suggestion\cite{shimano14,aoki_prb15} that third-harmonic generation (THG) in a superconductor can be attributed to  the resonant excitation of collective amplitude (Higgs) fluctuations of the SC order parameter, it has been recently shown\cite{cea_prb16,cea_raman_prb16} that the THG signal is dominated by lattice-modulated charge fluctuations (LCF). The basic argument is actually very simple. 
On general grounds\cite{Mahan}, the  average current $\bJ$ of  a system of electrons   in the presence of the e.m. gauge field $\bA$ is composed by two contributions%
\be
\lb{jgen}
J_\alpha=\langle j_\alpha\rangle-{\langle \rho_{\a\b} \rangle e^2}A_\beta, \quad \rho_{\a\b}\sim\sum_\bk (\pd^2_{\a\b}\e_\bk)\rho_\bk
\ee
where   $\bj$ is the paramagnetic current and  $\rho_{\a\b}$ is the diamagnetic tensor, which corresponds to the density operator $\rho_\bk$ times derivatives $\pd^2_{\a\b}\equiv \pd_{k_\alpha}\pd_{k_\b}$ of the band dispersion $\e_\bk$\cite{scalapino,review}.  In linear-response theory\cite{Mahan} one retains only terms linear in $\bA$, $J^L_\a\sim K_{\a\b}^L A_\b$. The paramagnetic contribution is then proportional to the current-current response function $ \langle j_\a\rangle\sim \langle j_\a  j_\b \rangle A_\b$, while in the diamagnetic term, that is already linear in $\bA$, one just replaces the density  with its average value, so that  $\langle \rho_{\a\b}\rangle$ scales with $n/m^*$, $m^*$ being the effective electronic mass.  If one is interested in the non-linear optical response the averages in Eq.\ \pref{jgen} should be computed to next order in $\bA$, so one is left with the correlation function measuring  density fluctuations modulated by the derivative of the band dispersion, i.e. LCF:
\be
\lb{jnlgen}
J^{NL}_{\a}=K^{NL}_{\a\b,\g\d}A_\beta A_\g A_\d, \quad K^{NL}_{\a\b;\g\d}\sim \langle \rho_{\a\b} \rho_{\g\d} \rangle
\ee
where Eq.\ \pref{jnlgen} has to be considered a convolution in  time and space. Apart from the modulation factors due to the derivative of the band dispersion, the non-linear response $K^{NL}(\omega)$ of Eq.\ \pref{jnlgen} probes density-like fluctuations, that in the SC state diverge for a frequency $\omega$ equal to the threshold $2\Delta_0$ above which Cooper pairs (CP) proliferate. In a typical non-resonant Raman experiment such a divergence is seen when the difference $\omega\equiv \omega_{in}-\omega_{out}$ between the incident ($\omega_{in}$) and scattered ($\omega_{out}$) light matches the $2\Delta_0$ value\cite{hackl_review}. In transmission experiments an incident monochromatic field oscillating at frequency $\omega$ generates a non-linear current \pref{jnlgen} oscillating at $3\omega$, with an amplitude $K^{NL}(2\omega)$ that is resonantly enhanced when 
the frequency $2\omega$ of the incoming $\bA^2$ field coincides with the $2\Delta_0$ value where LCF are peaked. 

In addition to this effect, there can be a subleading contribution coming from the amplitude (Higgs) fluctuations of the SC order parameter. 
From the technical point of view, this contribution appears as a vertex correction of the LCF response function in the pairing channel\cite{aoki_prb15,cea_prb16}. It then corresponds to accounting for all the intermediate virtual processes which convert the particle-hole excitation created by the incoming field in a pair fluctuation. However, this contribution is orders of magnitude smaller than the one due to the LCF alone\cite{cea_prb16}, since in a BCS superconductor the density-like fluctuations are decoupled from the Higgs mode\cite{varma_prb82,varma_review,cea_prl15}. In the strong-coupling limit the Higgs corrections become more relevant\cite{aoki_sc_prb16}, due however to broadening effects that wash out also the sharp resonance at $2\Delta$ found in the BCS limit and observed experimentally\cite{shimano14}.

As it is evident from Eq.\ \pref{jnlgen}, the non-linear response admits in general a non-trivial  dependence on the polarization of the incoming  e.m. field. In the two-dimensional one-band model considered  in Ref. \cite{cea_prb16} it has been predicted that the THG can vary by orders of magnitude by changing the relative direction between the e.m field and the axes of the lattice. However, in this paper we show that the strength of this effect strongly depends on the band structure and on the form of the pairing interaction. By computing the THG within a multiband model for the prototype NbN superconductor we show that the polarization dependence of the signal is strongly suppressed,  challenging its experimental observation. As far as the LCF contribution is concerned we show that to correctly compute the polarization dependence of the THG one must include the effect of the  instantaneous non-linear electronic response. This term, neglected in the recent analysis of Ref.\ \cite{shimano17}, does not influence the singular behavior of the non-linear response functions at $\omega=\Delta_0$, 
but it suppresses the polarization dependence of the THG, that is further smeared out by a realistic simulation of the finite-width of the pulse. For what concerns the Higgs signal we show that it remains subleading and its polarization dependence depends in general on the form of the pairing interaction. These results suggest that  the isotropy of the THG signal recently reported in Ref.\ \cite{shimano17} could be completely recovered once that realistic smearing effects on the LCF response due to disorder are included.

\section{Derivation of the non-linear response}
The starting model is a multiband generalization of Ref.\ \cite{cea_prb16}:%
\bea
H&=&\sum_{\mathbf{k},\sigma,a}\xi^a_\mathbf{k}c^\dagger_{\mathbf{k}\sigma,a}c_{\mathbf{k}\sigma,a}-\frac{1}{N_s} \sum_{\bq,ab} U_{ab}\Phi^\dagger_{\D,a}(\bq)\Phi_{\D,b}(\bq)\nn\\
\lb{hmodel}
&+&\frac{1}{2}\sum_\bq V(\bq)\Phi^\dagger_{\rho}(\bq)\Phi_{\rho}(\bq)
\eea
where $\xi_{\bk}^a=\e_\bk^a-\mu$ is the band dispersion in each $a$ band with respect to the chemical potential $\mu$, $U_{ab}$ is the matrix of the SC couplings, $\Phi_{\Delta,a}(\bq)=\sum_{\bk}c_{-\mathbf{k}+\mathbf{q}/2\downarrow,a}c_{\mathbf{k}+\mathbf{q}/2\uparrow,a}$ is the pairing operator for each band, $V(\bq)$ is the Coulomb potential and $\Phi_\rho(\bq)=\sum_{\bk,a}c^\dagger_{\mathbf{k}+\bq \sigma,a}c_{\mathbf{k}\sigma,a}$ is the total density operator.   The band dispersion for NbN follows from a tight-binding fit within the manifold of $xy,xz,yz$ $d$ orbitals on the fcc lattice, as suggested in Ref.\ \cite{shimano17}. By assuming only intra-orbital hopping one has:
\bea
\e^{xy}(k_x,k_y,k_z)&=&4t\cos\frac{k_x}{2}\frac{k_y}{2}+2t'(\cos k_x+\cos k_y)+\nn\\
\lb{exy}
&+&4 t"\left(\cos\frac{k_x}{2}\cos\frac{k_z}{2}+\cos\frac{k_y}{2}\cos\frac{k_z}{2}\right),
\eea
\bea
\lb{exz}
\e^{xz}(k_x,k_y,k_z)&=&\e^{xy}(k_x,k_z,k_y), \\
\lb{eyz}
\e^{yz}(k_x,k_y,k_z)&=&\e^{xy}(k_y,k_z,k_x)
\eea
which are obtained by ciclic permutations of the wavevector indexes between the bands. To make the derivation simpler we will first discuss the case where pairing has only {\em intraband} character, so that $U_{ab}=U\delta_{ab}$, and we will discuss later on the consequences of a more general pairing interaction.

 The general strategy to compute the non-linear response has been outlined in Ref.\ \cite{cea_prb16}: by means of the Hubbard-Stratonovich transformation one decouples the interaction terms of the model \pref{hmodel} and derives an effective action written in terms of the collective  charge ($\rho$), SC phase ($\theta^a$) and SC amplitude ($\Delta^a$) collective fluctuations. By adding also the gauge field $\bA$ by means of the minimal-coupling Peierls substitution, one can obtain the effective action $S[A]$ for the gauge field $\bA$ up to the fourth order, by retaining the coupling between $\bA$ and the collective degrees of freedom. 
  {As detailed in the Appendix A}, $S[A]$ can then be obtained in the $\bq\ra 0$ long-wavelength limit as  straightforward extension of the result of Ref. \cite{cea_prb16}:
 \bea
S[A]&=&\frac{1}{2}\sum_{a}  e^4A_\a A_\b \chi^a_{\a\b,\g\d}A_\g A_\d+
2e^2A_\a^2\chi^a_{A^2_\a\D}\Delta^a\nn\\
&+&2ie^2A_\a^2 \chi^a_{A^2_\a \rho}\left[\rho +(i\Omega_n \theta^a/2)\right]+\nn\\
\lb{SA}
&+&S[\rho,\theta^a,\Delta^a],
\eea 
where the explicit dependence of each term on the $i\Omega_n$ bosonic Matsubara frequency has been omitted for simplicity. The last term of Eq.\ \pref{SA} describes the collective fluctuations of the total density and of the SC amplitude and phase in each band. In the presence of Coulomb interactions and for intra-band pairing only they have a very simple form in the long wavelength limit\cite{cea_prb16,cea_raman_prb16}:
\be
\lb{Sr}
S[\rho,\theta^a,\Delta^a]=\frac{1}{2} \sum_a -\chi^a_{\rho\rho} \left|\rho +(i\Omega_n \theta^a/2)\right|^2+X_{\D\D}^a \Delta_a^2,
\ee
where $X^a_{\D\D}=(4\Delta_0^2-(i\Omega_n)^2)\sum_{\bk}F^a_\bk(i\O_n)$ denotes the inverse amplitude-mode propagator, and we defined the response functions 
\begin{subequations}
\bea
\lb{chi_0}
\chi^a_{\alpha\beta;\g\d}&=&\langle \rho^a_{\a\b} \rho^a_{\g\d}\rangle+n^{el}_{\a\b\g\d}\\
\lb{chi_0nonloc}
\langle \rho^a_{\a\b} \rho^a_{\g\d}\rangle&=&{\Delta_0^2}\sum_{\mathbf{k}}\pd^2_{\a\b}\e^a_\bk\pd^2_{\g\d}\e^a_\bk F^a_\bk(i\O_n)\\
\lb{chi_0loc}
n^{el}_{\a\b\g\d}&=&
\sum_\bk\frac{\partial^4_{\alpha\beta\gamma\delta}\e^a_\bk}{12N_s} \left[1- \frac{\xi^a_\bk\tanh(E^a_\mathbf{k}/2T)}{E^a_\mathbf{k}}\right]
\eea
\end{subequations}
\bea
\lb{chi_mix}
\chi^a_{A^2_\alpha\rho}&=&\langle \rho^a_{\a\a} \rho^a\rangle={\Delta_0^2}\sum_{\mathbf{k}}(\pd^2_{\a\a}\e^a_\bk)  F^a_\bk(i\O_n) \\
\lb{chi_rho}
\chi^a_{\rho\rho}&=&\langle \rho^a  \rho^a\rangle={\Delta_0^2}\sum_{\mathbf{k}} F^a_\bk(i\O_n) \\
\lb{chi_D}
\chi^a_{A^2_\alpha\Delta}&=&\langle \rho^a_{\a\a} \Delta^a\rangle ={\Delta_0}\sum_{\mathbf{k}}(\pd^2_{\a\a}\e^a_\bk) \xi^a_{\bk} F^a_\bk(i\O_n)
\eea
where $F^a_\bk(i\O_n)=\frac{1}{N_s} \frac{\tanh(E^a_\mathbf{k}/2T)}{E^a_\mathbf{k}\left[(i\Omega_n)^2-4(E^a_\mathbf{k})^2\right]}$ and $E^a_{\bk}=\sqrt{(\xi^a_\bk)^2+\Delta_0^2}$. The second term of Eq.\ \pref{chi_0}, defined by Eq.\ \pref{chi_0loc}, is  constant in frequency and it gives rise  to a  contribution local in time in the action \pref{SA}, accounting for the instantaneous electronic response in the current: $J^{el}_\a(t)\sim n^{el}_{\a\b\g\d} A_\b(t)A_\g(t)A_\d(t)$. {This term originates from the fact that in a lattice model the minimal coupling to a constant gauge field $\bA$ amounts to replacing the wavevector $\bk$ with $\bk+\bA$ in the band dispersion $\e_\bk$. As a consequence, the bare current obtained as a derivative of the Hamiltonian with respect to $\bA$ contains all orders in $\bA$, leading to this additional instantaneous contribution (see Appendix A for further details). This term has been neglected in previous work\cite{cea_prb16,shimano17} since id does not contribute to   singular behavior of the non-linear response function\cite{cea_prb16}, responsible for the enhancement of the THG at the resonance condition $\omega=\Delta_0$. However, it contributes to the polarization dependence, so it cannot be ignored once that a quantitative estimate of this effect is required. On the other hand, whenever the system is in the low-density limit, where the band dispersion can be approximated by a parabola, this contribution becomes progressively irrelevant, being proportional to a fourth-order derivative of the band dispersion. }
The $\chi^a_{\a\b,\g\d}$ represents the "bare" LCF response, which is dressed by the fluctuations in the phase/density and amplitude sectors, due to the couplings to these fields in the second line of Eq.\ \pref{SA}. {We notice that additional diagrams entering in the bare response $\chi^a_{\a\b\g\d}$ and not proportional to density fluctuations vanish in the SC state at long wavelengths (see Appendix A). }
The full response can be derived from Eq.\ \pref{SA} by Gaussian integration of the collective electronic excitations, which is equivalent to adding vertex corrections in the particle-hole and particle-particle channels.

 \begin{figure}[t]
\includegraphics[width=8cm,clip=true]{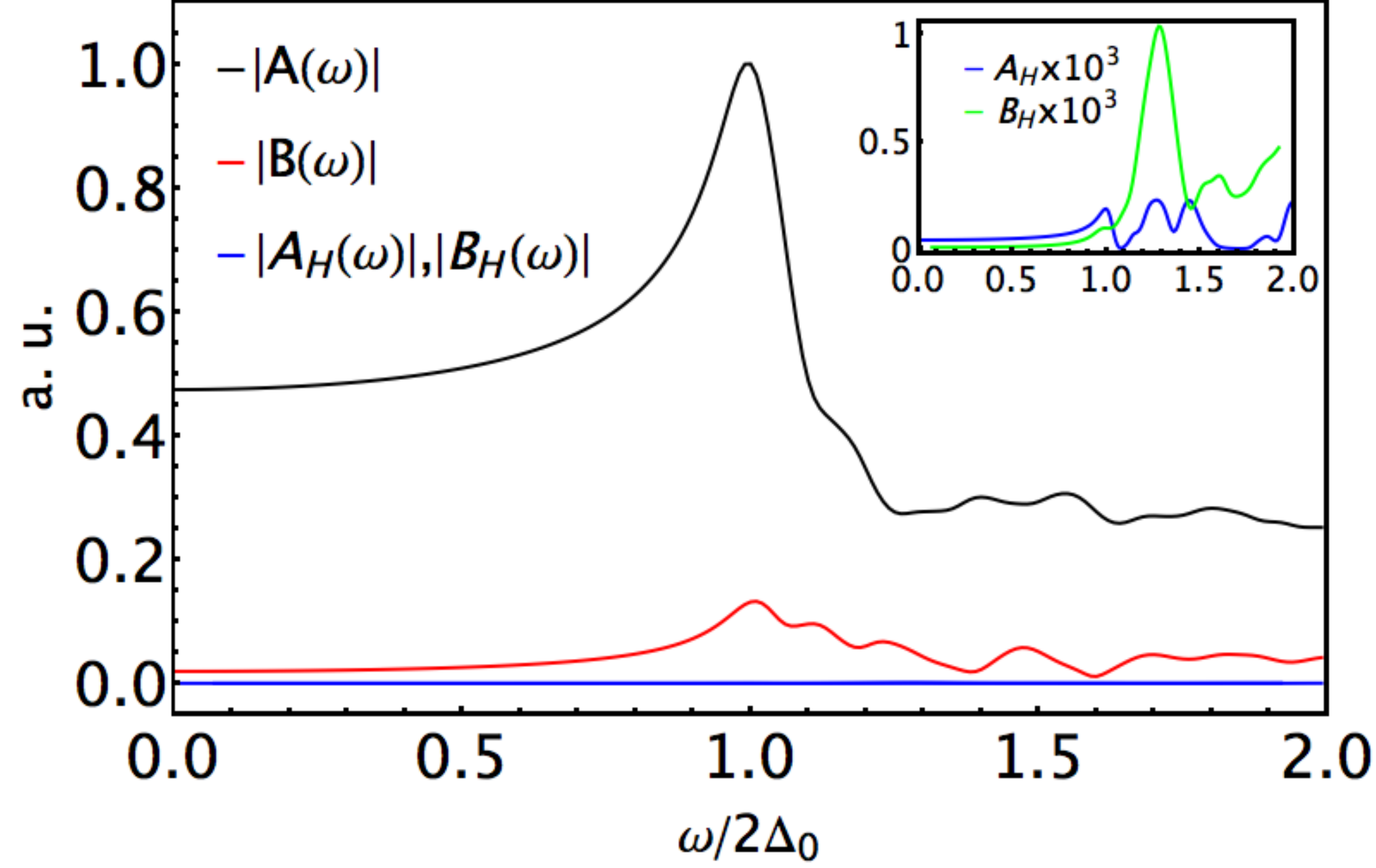}
\caption{Frequency dependence of the various contributions \pref{alcf}-\pref{blcf} and \pref{ahiggs}-\pref{bhiggs} to the non-linear current. The calculations have been done with the band structure of Eq.\ \pref{exy}-\pref{eyz}, with the same parameter values used in Ref.\ \cite{shimano17}, i.e. $t=-0.72$ eV, $t'=-0.15$ eV, $t"=0.12$ eV, $\mu=-0.6$ eV. The coupling $U=0.27$ eV is chosen to match the experimental $T=0$ value $\Delta_0=0.65$ THz.  Inset: expanded view of the Higgs contributions only.}
\label{aom}
\end{figure}

\section{Computation of the third-harmonic intensity}
For the band dispersion \pref{exy}-\pref{eyz} and a field applied in the $xy$ plane, as in the geometrical configuration of Ref.\ \cite{shimano17}, only the terms $\chi^a_{xx;xx}=\chi^a_{yy;yy}$ and $\chi^a_{xy;xy}$ survive in the  $A^4$ term of Eq.\ \pref{SA}. Let us first compute the vertex corrections in the phase/density channel. By gauging away the total density, as explained for the multiband case in Ref.\ \cite{cea_raman_prb16}, and performing the Gaussian integration over the $\theta^a$ fields, one easily finds:
\bea
S&=& \frac{e^4}{2}\int dt dt' \left\{A_x^2(t)C_{xx}(t-t')A_x^2(t')\right.+\nn\\
&+&A_y^2(t)C_{yy}(t-t')A_y^2(t')+\nn\\
&+&[A_x^2(t)A_y^2(t')+A_x^2(t)A_y^2(t')]C_{xy}(t-t')+\nn\\
\lb{stot}
&+&\left.4\left[A_x(t)A_y(t)A_x(t')A_y(t')\right]D_{xy}(t-t')\right\}
\eea
where we defined
\bea
\lb{cxx}
C_{xx}&=&\sum_a \chi^a_{xx;xx}-\frac{(\chi^a_{A^2_x\rho})^2}{\chi^a_{\rho\rho}}\\
\lb{cxy}
C_{xy}&=&\sum_a \chi^a_{xx;yy}-\frac{\chi^a_{A^2_x\rho}\chi^a_{A^2_y\rho}}{\chi^a_{\rho\rho}}\\
\lb{dxy}
D_{xy}&=&\sum_a \chi^a_{xy;xy}
\eea
and analogous expression for $C_{yy}$. The exact form of the vertex corrections in the particle-hole channel, i.e. the second terms in Eq.\ \pref{cxx} and \pref{cxy}, depend on the pairing interaction $U_{ab}$. When also interband interactions are present the phase sector admits massive Leggett modes, making the computation more involved. The result, derived explicitly in the two-band case in Ref.\ \cite{cea_raman_prb16}, show that also in this case vertex corrections retain a polarization dependence. The only case where the phase/density corrections are polarization independent is the unrealistic situation where $U_{ab}=U$, i.e. {\em intraband} pairing interactions equal the {\em interband} ones. Indeed in this case, considered in Ref.\ \cite{shimano17}, one can define a single collective phase/density field, removing the polarization dependence of the vertex corrections.

The non-linear current $J_\alpha^{NL}$  is easily found by functional derivative with respect to $A_\alpha(t)$ in the action \pref{stot}.
For  a monocromatic incident field $\bA=\bar \bA\cos(\Omega t)$  there is a component of the current oscillating at three times the incident frequency, with an amplitude controlled by the non-linear kernel evaluated at $2\Omega$. The THG is a measure of the transmitted electric field $\bE^{tr}$, which is proportional to the current, so that  $I^{THG}_{\alpha}(\Omega)\propto \left|	\int\,dt J^{NL}_{\alpha}(t)e^{3i\Omega t} \right |^2$. For a field $\bar\bA$ applied at a generic angle $\theta$ in the $xy$ plane, as in the configuration of Ref.\ \cite{shimano17}, the current can be decomposed in a component parallel $J^{THG}_\pll(\Omega,\theta)$ and perpendicular $J^{THG}_\perp(\Omega,\theta)$ to $\bar \bA$. With straightforward algebra one derives from Eq.\ \pref{stot} that\cite{shimano17}
\bea
\lb{jpll}
J^{THG}_\pll(\Omega,\theta)&=&A(2\Omega)+2B(2\Omega)\sin^2 2\theta,\\
\lb{jperp}
J^{THG}_\perp(\Omega,\theta)&=&B(2\Omega)\sin 4\theta
\eea
where
\bea
\lb{alcf}
A(\omega)&=&C_{xx}(\omega)\\
\lb{blcf}
B(\omega)&=&\frac{1}{4}(C_{xy}+2D_{xy}-C_{xx})
\eea
where we used the fact that $C_{xx}=C_{yy}$ after summation over momenta and band indexes in Eq.\ \pref{cxx}. 

The same arguments hold also for the vertex corrections in the amplitude  channel, i.e. for the Higgs contribution. It  can be  derived with the same procedure, i.e. by Gaussian elimination of the $\Delta^a$ fields in Eq.\ \pref{SA}, so that Eq.s\ \pref{cxx} and \pref{cxy} acquire two new terms:
\bea
\lb{cxxh}
C^H_{xx}&=-&\sum_a \frac{(\chi^a_{A^2_x\D})^2}{X^a_{\D\D}}\\
\lb{cxyh}
C^H_{xy}&=&-\sum_a \frac{\chi^a_{A^2_x\D}\chi^a_{A^2_y\D}}{X^a_{\D\D}}
\eea
As a consequence also the Higgs contribution to the non-linear current admits the decomposition \pref{jpll}-\pref{jperp}, with 
\bea
\lb{ahiggs}
A^H&=&C^H_{xx}(\omega)\\
\lb{bhiggs}
B^H&=&\frac{1}{4}(C^H_{xy}-C^H_{xx})
\eea
Once more, $B^H$ for a generic pairing interaction is not zero. The vanishing of $B^H$ in  Ref.\ \cite{shimano17} is  due to the specific choice of an interband pairing identical to the intraband one, which is the {\em only} case where the amplitude fluctuations collapse in a single effective Higgs field.  

The relative magnitude of the various $A,B,A^H,B^H$ terms is shown in Fig.\ \pref{aom}. The Higgs terms $A^H, B^H$ are largely subdominant with respect to the $A,B$ ones, due to the particle-hole symmetry of the SC ground state, which suppresses the $\chi_{A^2_x\Delta}^a$ susceptibilities \cite{varma_prb82,varma_review,cea_prl15}. It is worth noting that in the present case of almost half-filled bands also the vertex corrections in the particle-hole channel, i.e. the second terms in Eq.s\ \pref{cxx}-\pref{cxy},  are quantitatively irrelevant. Thus the present computation of the LCF response is quantitatively robust with respect to variations in the form of the pairing interactions.

\begin{figure}[t]
\includegraphics[width=8cm,clip=true]{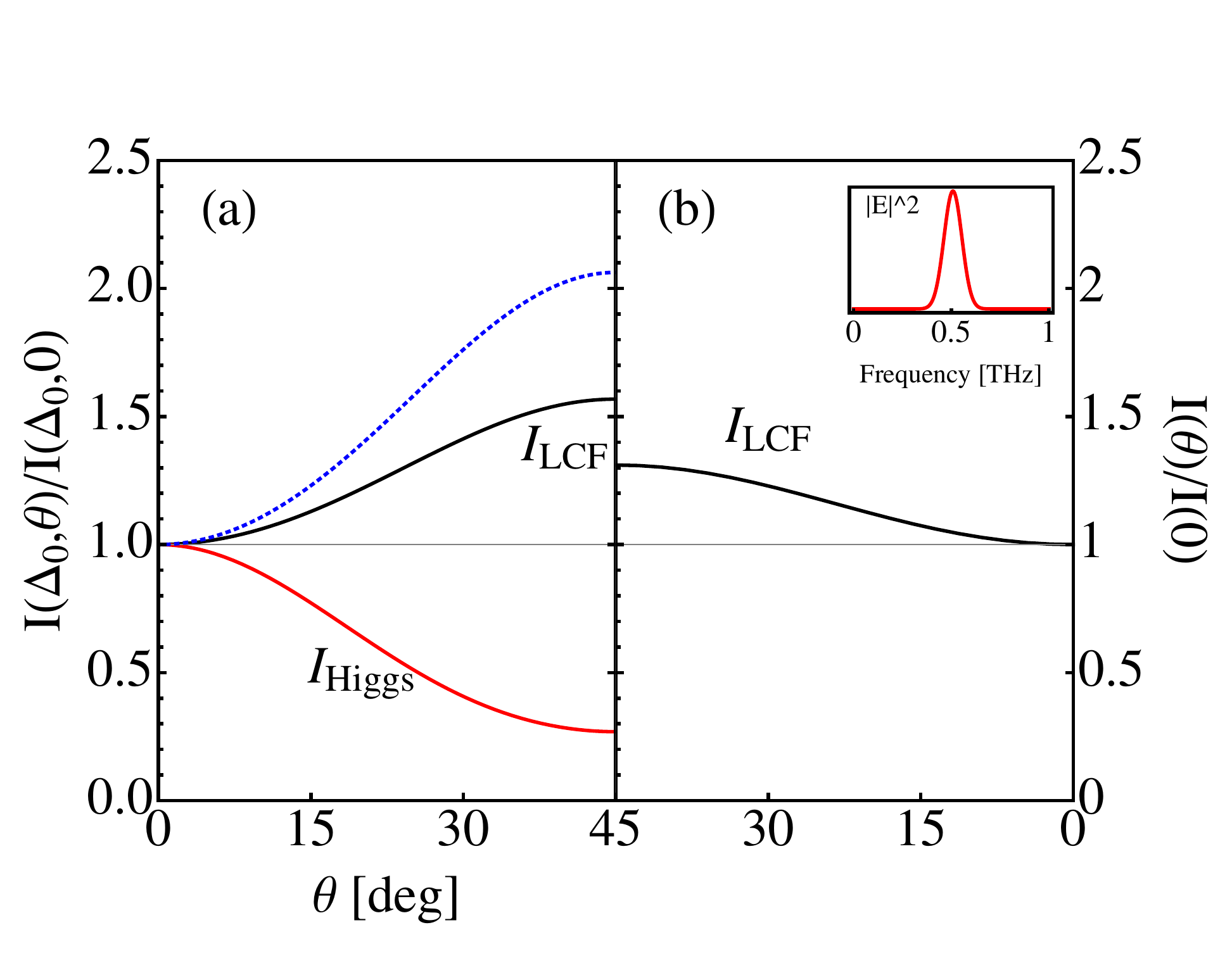}
\caption{(a) Relative angular variation of the THG intensity for the model \pref{hmodel}.  LCF contribution (solid black curve) and Higgs contribution (normalized to $I^H(\Delta_0,\theta=0)$) alone (red curve). The dotted black line is the result of Ref.\ \cite{shimano17}. (b) Angular variation of the LCF contribution for an incident electric field simulating the experimental situation of Ref.\ \cite{shimano17}, as shown in the inset. }
\label{angle}
\end{figure}

The strong resonance of LCF contribution at twice the gap value in Fig.\ \ref{aom} explains the enhancement of the non-linear current \pref{jpll}-\pref{jperp}  at $\Omega=\Delta_0$.  A first estimate of the angular dependence of the THG intensity at resonance in the direction of the applied field, as measured in Ref.\ \cite{shimano17}, can then be obtained as 
\be
I(\Omega=\Delta_0,\theta)\propto |J^{THG}_\pll(\Omega=\Delta_0,\theta)|^2.
\ee
The relative angular variation of the LCF contribution for the model \pref{hmodel} is shown in Fig.\ \ref{angle}a (continue black curve). As one can see, with respect to the two-dimensional toy-model considered in Ref.\ \cite{cea_prb16}, the angular variations are strongly suppressed for the three-dimensional band structure of NbN. For the sake of completeness, we also show  the largely subdominant Higgs contribution  alone (red curve), that displays an even stronger angular dependence. The softening of the angular variation of the LFC part  with respect to the result of Ref.\ \cite{shimano17}, represented by the dotted blue line in Fig.\ \ref{angle}a, is due to the 
constant term $n^{el}_{\a\b\g\d}$ in Eq. \pref{chi_0}. Indeed this term, neglected in Ref.\ \cite{shimano17},  reduces the ratio $B/A$ at resonance,  and even more away from it, see Fig.\ \pref{aom}. This effect further reduces the observable polarization dependence when one considers the more realistic case of an incident electric field with a finite spectral width. Simulating the experimental pump used in Ref.\ \cite{shimano17}  as 
$\mathbf{A}(t)=\bar{\mathbf{A}}F(t)$, where $F(t)=e^{-\left[t\sigma/ (4\sqrt{\ln2})\right]^2}\cos\left(\O t\right)$, we can compute the time-dependent non-linear current as $J_\pll(t,\theta)=-F(t)\int dt' K_\pll(\theta,t-t')F^2(t')$  where $K_\pll(\theta,t-t')$ is the non-linear kernel corresponding to Eq.\ \pref{jpll}. In the experimental configuration of Ref.\ \cite{shimano17} the wavepacket has a central frequency $\Omega=0.5$THz and width $\sigma\simeq0.15$THz, so that the experimental signal is integrated 
in a range  1.3 THz-1.7 THz centred around the third-harmonic frequency $3\Omega=1.5$THz. By performing the same procedure for our model we obtain the result shown in Fig.\ \ref{angle}b, where the angular dependence of the THG is further smeared out, with a relative enhancement of the intensity of $I(45°)/I(0°)\sim 1.3$. 

\section{Discussion and conclusions}

The softening of the relative enhancement of the THG intensity shows in Fig.\ \ref{angle}b is a direct consequence of the broadening of the $2\Delta_0$ resonance of the non-linear response when one accounts for the experimental configuration.  This example suggests the ratio $B/A$ may be also suppressed by disorder effects, that has been shown to smear out considerably the SC Higgs resonance within realistic microscopic models for disorder\cite{cea_prl15}. As a consequence, to fully capture the isotropy of the experimental THG signal  reported in Ref.\ \cite{shimano17} the scattering by defects could play a relevant role. 

In general, our results demonstrate that while the predominance of charge fluctuations over the Higgs contribution is a generic feature also in multiband systems, since it is based on the weak coupling of the Higgs mode to the density in BCS superconductors\cite{varma_prb82,varma_review,cea_prl15,cea_prb16,cea_raman_prb16}, an exact  quantitative estimate of the polarization dependence of the THG  is strongly model-dependent. As a consequence, any modification on the description of the band structure can lead to quantitative  change on the THG polarization dependence, even though the basic underlying mechanism is the enhancement of charge fluctuations in the SC state, as proposed in the present work. For example, in the specific case of NbN considered here an estimate of the Slater-Koster matrix elements  shows that  a-priori a tight-binding model based on the $d$ orbitals on the fcc lattice should include also inter-orbital hopping terms, neglected in the model \pref{hmodel}. This fact can have direct consequences on the definition of non-linear response function after the minimal-coupling Peierls substitution, and then on the polarization dependence of the THG signal due to charge fluctuations. For the same reason, it is hard to predict how the Higgs contribution will change in the strong-coupling limit of the Hubbard-Holstein model recently considered in Ref.\ \cite{aoki_prb15} within the context of single-band superconductors. Indeed, as  discussed in more details in Appendix A, the processes making the Higgs visible in this limit have a full tensorial structure, so one does {\em not} expect them to be polarization independent. Thus a precise quantification of the relevant processes at strong disorder and/or interaction remains an interesting problem for future work.

In summary, we computed the THG in a multiband model appropriate for NbN. We have shown that the Higgs contribution to the THG signal remains negligible, and it is in general polarization dependent. The isotropy of the Higgs contribution recently claimed in Ref.\ \cite{shimano17} is a peculiarity of the case where interband pairing interactions coincide with the intraband ones, which is far from being a general feature of SC multiband systems. As far as the dominant charge fluctuations are concerned, we have shown that the instantaneous electronic response and the finite spectral width of the pump contribute to suppresses the polarization dependence of the THG, challenging its experimental detection in realistic experimental situations in disordered films. 



\acknowledgements
  
We acknowledge useful discussions with J. Lorenzana. This work has been supported  by Italian MAECI under the Italian-India
collaborative  project  SUPERTOP-PGR04879 and by the Graphene flagship.

\appendix
\section{Derivation of the effective action}

The derivation of the effective action \pref{SA} follows the same steps outlined in Ref.\ \cite{cea_prb16} for the single-band case and extended in Ref.\ \cite{cea_raman_prb16} to the two-band case. The starting point is the introduction of a set of bosonic complex fields $\psi^a_\D(\tau)$ which decouple the pairing term in Eq.\ \pref{hmodel}.  At $T<T_c$ one can choose to represent the SC fluctuations  in polar (amplitude and phase)  coordinates, by decomposing $\psi^a _\D(\tau)=[\Delta^a_0 +\Delta^a(\tau)]e^{i\theta^a(\tau)}$, where $\Delta^a(\tau)$ represents the amplitude fluctuations of $\psi^a_\D$ around the mean-field value $\Delta^a_0$ of the SC order parameter in the band $a$ and $\theta^a$ its phase fluctuations. By making a Gauge transformation $c_{i,a}\ra c_{i,a}e^{i\theta^a/2}$ the dependence on the phase degrees of freedom is made explicit in the action. Analogously, the last line of Eq.\ \pref{hmodel} is decoupled by introducing a HS field $\psi_\rho=\rho_0+\rho$, which couples to the total electronic density $\Phi_{\rho}$ and represents the density fluctuations $\rho$ of the system around the mean-field value $\rho_0$. Finally, the Gauge field $\bA$ can be introduced by means of the Peierls substitution  $c^\dagger_{i+\hat x,a}c_{i,a}\rightarrow c^\dagger_{i+\hat x,a}c_{i,a}e^{ie \bA\cdot \hat x}$, that modifies only the kinetic part of the Hamiltonian, leading to the shift $\e_\bk\ra\e_{\bk+\bA}$ in the band dispersion.

After the Hubbard-Stratonovich decoupling the action is quadratic in the fermionic fields, so that one can integrate them out leading to the effective action for the collective bosonic fields only. The equilibrium values of the HS field appear in the mean-field action $S_{MF}$,  
\begin{equation}
\lb{smf}
S_{MF}=\frac{N}{T}\sum_{ab}\Delta_0^aU_{ab}^{-1}\Delta_0^b-\text{Tr}\sum_a\ln(-G_{0,a}^{-1})
\ee
where   $G_{0,a}^{-1}=i\o_n\hat\s_0-\xi^a_\bk\hat\s_3+\D_0^a\hat\s_1$ is the inverse BCS Green's function for the electrons in the $a$ band and $\hat\sigma_i$ are Pauli matrices. The minimization of $S_{MF}$ with respect to $\Delta^a_0$ gives the usual self-consistent mean-field equations for the SC gap. In the case of diagonal pairing matrix $U_{ab}=U\delta_{ab}$ the BCS order parameter $\Delta_0^a\equiv \Delta_0$ is the same in all the bands. 
By adding SC and density fluctuations one obtains the effective action of collective modes as an expansion in powers of the HS fields:
\begin{equation}
\lb{seff}
	S_{eff}[\Delta^a,\theta^a,\rho,\bA]=S_{MF}+S_{FL}[\Delta,\theta,\rho,\bA]\quad,
\end{equation}
where  
\be 
\lb{sfl}
S_{FL}=\sum_{n\ge1,a}\frac{\text{Tr}(G^a_0\Sigma^a)^n}{n}
\ee
is the fluctuating action, with the trace acting both in spin and momentum space. Here $\Sigma^a_{kk'}$ denotes the self-energy for the fluctuating fields, which reads explicitly:
\begin{widetext}
\bea
\Sigma^a_{kk'}&=&
-\sqrt{\frac{T}{N}}\Delta^a(k-k')\sigma_1-\sqrt{\frac{T}{N}}\rho(k-k')\sigma_3-
\sqrt{\frac{T}{N}}\frac{i}{2}\theta^a(k-k')\left[
(k-k')_0\sigma_3-(\xi^a_\bk-\xi^a_{\bk'})\sigma_0
\right]-\nn\\
&-&\frac{T}{2N}\sum_{q_\a,\a}\theta^a(q_1)\theta^a(q_2) \frac{\pd^2 \xi^a_\bk}{\pd k^2_\a} \sin\frac{\bq_{1,\a}}{2}\sin\frac{\bq_{2,\a}}{2}
\sigma_3 \delta(q_1+q_2-k+k')+\nn\\
&+& A_\a(\omega-\omega')\frac{\pd \xi^a_\bk}{\pd k_\a}\sigma_0+\frac{1}{2} [A_\a A_\beta] (\omega-\omega')\frac{\pd^2 \xi^a_\bk}{\pd k_\a \pd k_\beta}\sigma_3+
\frac{1}{3!}[A_\a A_\beta A_\delta](\omega-\omega)\frac{\pd^3 \xi^a_\bk}{\pd k_\a \pd k_\beta \pd k_\gamma}\sigma_0+\nn\\
\lb{SELF_ENERGY}
&+&\frac{1}{4!}[A_\a A_\beta A_\gamma A_\delta ](\omega-\omega')\frac{\pd^4 \xi_\bk}{\pd k_\a \pd k_\beta \pd k_\gamma\pd k_\delta}\sigma_3
\eea
\end{widetext}
with $k=(i\Omega_n,\mathbf{k})$ and $\Omega_n=2\pi Tn$ bosonic Matsubara frequencies, and $\a=x,y,z$ denoting spatial indexes. In Eq.\ \pref{SELF_ENERGY} the symbol $[A_\alpha \cdots A_\delta](\omega)]$ denotes the Fourier transform of the product of various field components taken at the same time, e.g. 
$[A_\a A_\beta](\omega)\equiv \int\,d\omega' A_\a(\omega-\omega')A_\beta(\omega')$ is the Fourier transform of $A_\a(t) A_\beta(t)$. 
\begin{figure}
\centering
\includegraphics[scale=1]{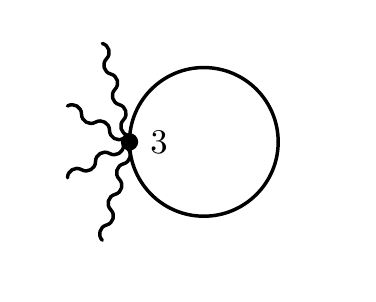}
\caption{
Diagrammatic representation of the instantaneous term \pref{chi_0loc}. Wavy lines denote the gauge field and solid line denote the Nambu Green's function. The number on the vertex denotes the insertion of the corresponding Pauli matrix.
}\label{fig-octopus}
\end{figure}

The second line of Eq.\ \pref{SELF_ENERGY} represents the transcription on the lattice of the usual $(\nb \theta)^2$ term for a continuum model, and analogously  the $[A_\alpha A_\beta] (\o)$ term that represents the transcription of the usual diamagnetic term $\bA^2 n/m$ in the continuum. { In addition, in contrast to the continuum model, the lattice self-energy \pref{SELF_ENERGY} depends in principle\cite{depalo_prb99,benfatto_prb04} on all higher-order powers of the $\theta$ and $\bA$ fields. 
In particular, the last term of Eq.\ \pref{SELF_ENERGY} is responsible for the new instantaneous term  $n^{el}_{\a\b\g\d}$ of Eq.\ \pref{chi_0loc}. Indeed, since $ \text{Tr}(G^a_0\sigma_3)\equiv n_\bk^a$ is simply the electron density in the $a$ band, one immediately recovers the instantaneous term defined in Eq.\ \pref{chi_0loc}, see also Fig.\ \ref{fig-octopus}.  }  Finally we observe that,  in contrast the the square 2D lattice considered in Ref.\ \cite{cea_prb16}, on the fcc lattice also a mixing of the various spatial components of the gauge fields $\bA$ is allowed at ${\cal O}(A^2)$ and beyond. 

\begin{figure}
\centering
\includegraphics[scale=0.45]{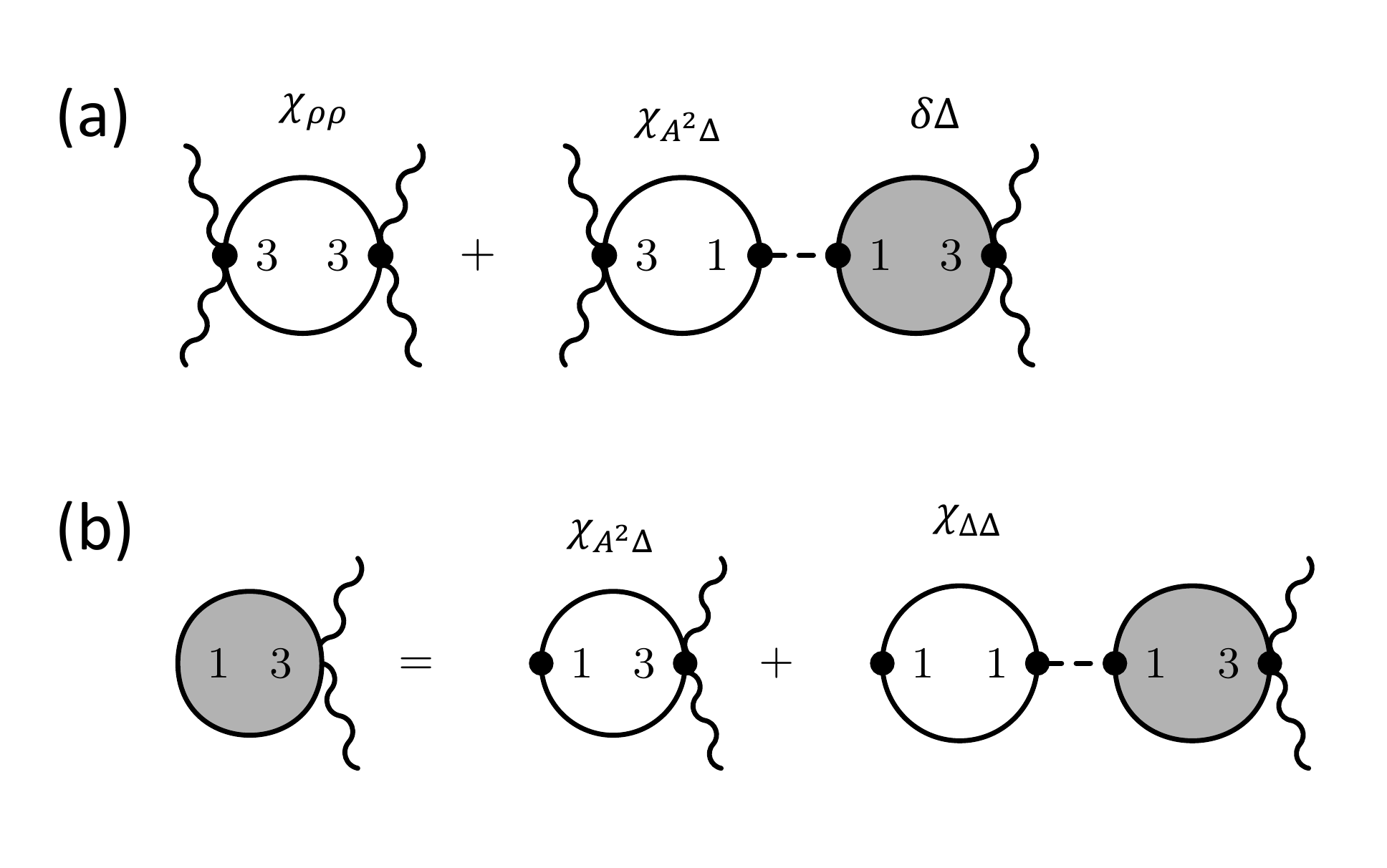}
\caption{
(a) Diagrammatic representation of the resonant terms of $S^{(4)}$, including vertex corrections in the Higgs channel. Here wavy lines denote the gauge field, solid lines the Nambu Green's function and the dashed line the pairing interaction. The shaded circle represents the variation of the order parameter due to the external perturbation, as defined in Eq.\ \pref{deltaDELTA}. (b) Vertex equation for $\delta\Delta$. Its solution leads to the RPA resummation of the $\chi_{\Delta\Delta}$ bubble, which defines the Higgs propagator $X^a_{\D\D}\equiv 1/U+\chi^a_{\D\D}$. The labels 1, 3 refer to the vertex insertions of the Pauli matrices $\hat \sigma_1$ and $\hat\sigma_3$, respectively. Here we omitted for simplicity the spatial indexes of the various fermionic susceptibilities.
}\label{Diagrams}
\end{figure}

Computing the trace in Eq.\ \pref{sfl} is equivalent to get an expansion in powers of the bosonic fields whose coefficients are fermionic susceptibilities obtained by mixing several $\hat\sigma_i$ Pauli 
matrices, establishing then a precise correspondence with the
different types of electronic excitations. More specifically,  $\hat \sigma_0$ insertion correspond to current-like fluctuations, $\hat \sigma_1$ to Higgs-like fluctuations and $\hat \sigma_3$ to density-like fluctuations, eventually modulated by derivatives of the band dispersion. This identification justifies the subscripts in Eq.\ \pref{chi_0}-\pref{chi_D}. 
The quadratic terms in $\Delta^a, \theta^a$ and $\rho$ define the spectrum of the collective modes, see Eq.\ \pref{Sr} above. As usual\cite{benfatto_prb04,cea_prl15,cea_prb16}, diagrams mixing two different Pauli matrices are subleading in the BCS limit. This implies for example that the coupling between the amplitude and phase/density modes, controlled by the fermionic susceptibility $\chi_{\rho\Delta}\sim \text{Tr}[G_0^a\sigma_3 G^a_0\sigma_1]$, can be neglected, as done in  Eq.\ \pref{Sr}. With lenghtly but straightforward calculations one can derive the effective action including also the gauge field, as given by Eq.\ \pref{SA} above. Here $X^a_{\D\D}\equiv 1/U+\chi^a_{\D\D}$ is the inverse Higgs propagator, obtained by RPA resummation of the amplitude susceptibility $\chi^ a_{\D\D}$ and using the self-consistence equation for the gap\cite{cea_prb16,cea_raman_prb16}. 

For the sake of simplicity we included only the leading diagrams responsible for the polarization dependence and the SC resonance. In particular Eq. \pref{chi_0} defines the most relevant term in the SC state, connected to lattice-modulated charge fluctuations. Integrating out the Higgs or the density/phase modes corresponds to add vertex corrections in the corresponding channels, as exemplified for a given band in the Higgs channel in Fig.\ \ref{Diagrams}. By denoting $\delta\Delta^a_{\a\b}$ the variation of the order parameter from its equilibrium value due to the external perturbation, it can be obtained by dressing $\chi^a_{A^2_\alpha \D}$ with the vertex correction in the amplitude channel (see fig. \ref{Diagrams}-(b)), so that:
\begin{equation}\lb{deltaDELTA}
\delta\Delta^a_{\a\b}(\omega)=e^2\frac{\chi^a_{A^2_\b\D}(\omega)}{X^a_{\D\D}(\omega)} A_\b^2(\o),
\end{equation}
that corresponds to Eq.s\ \pref{cxxh}-\pref{cxyh}.
{Notice that when the pairing matrix is assumed totally isotropic, $U_{ab}=U$, one can introduce a single Hubbard-Stratonovic field to decouple the pairing interaction. This implies that a single Higgs propagator $X_{\D\D}$ exists and the summation over band index in Eq.\ \pref{cxxh}-\pref{cxyh} leads to a vanishing of the $B^H$ term in Eq.\ \pref{bhiggs}, as indeed found in Ref.\ \cite{shimano17}. However, any other form of pairing interaction requires the introduction of separate Higgs fluctuations in each band, leading in general to an anisotropic contribution of the Higgs mode as well. 
Analogously for the particle/hole channel one adds fluctuations are the RPA level and obtains the action given by Eq.\ \pref{stot}. We notice that the present derivation is completely equivalent to the usual diagrammatic expansion. This issue has been recently discussed for the multiband case in Ref.s \cite{cea_raman_prb16} and \cite{maiti_prb17}, where the effective-action and diagrammatic expansion have been used, respectively, to derive the Raman response, leading to the same final result. }

\begin{figure}
\centering
\includegraphics[scale=0.4]{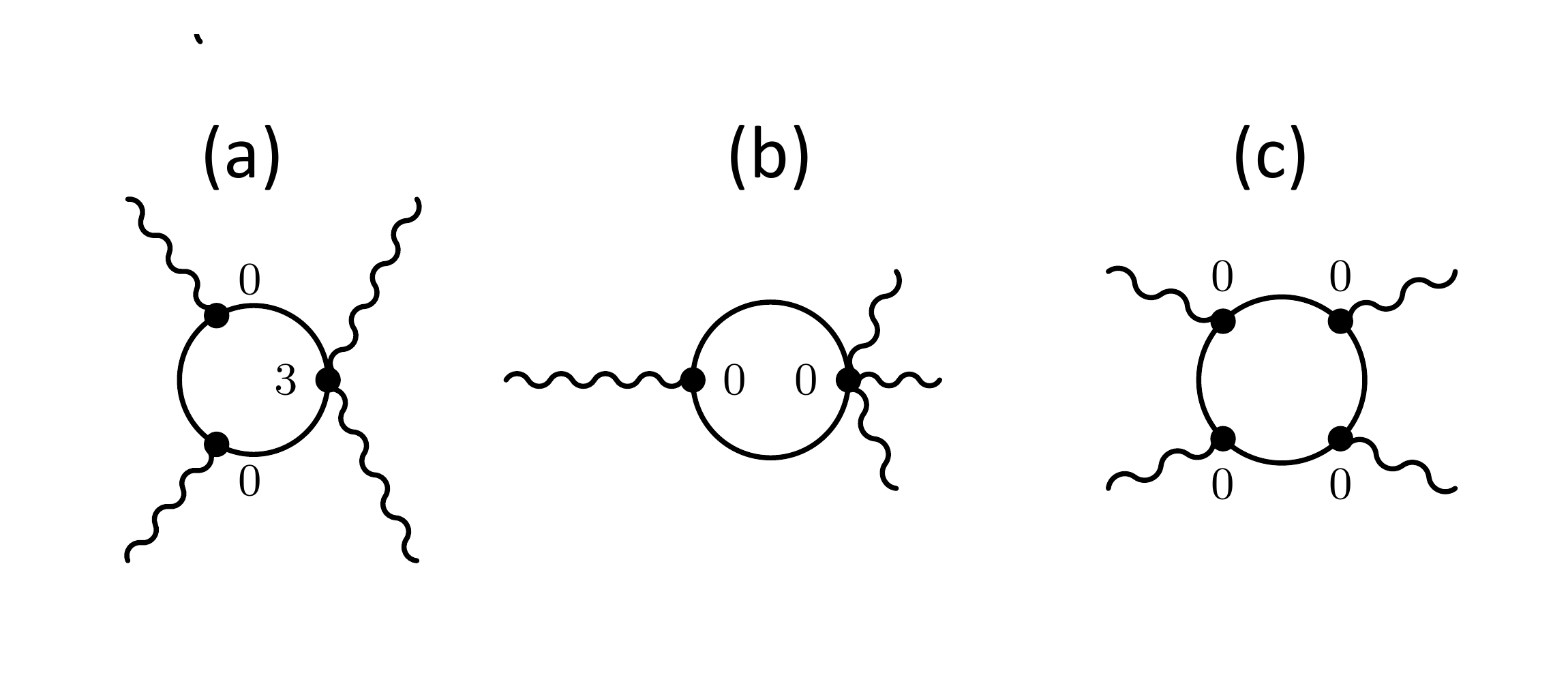}
\caption{
(a)-(c) Additional diagrams contributing to $S^{(4)}$. They vanish identically at $\bq=0$ as $T\ra 0$ in the BCS limit.
}\label{Diagrams2}
\end{figure}

In addition to the resonant diagrams and the instantaneous response included in Eq.\ \pref{SA}  one can have in principle several other terms of order $A_\a^4$, coming from the insertion of various $A_\a^n$ term of the self-energy \pref{SELF_ENERGY},  as shown in Fig.\  \ref{Diagrams2} (a)-(c). These terms can be defined as paramagnetic ones, since they all carry out a current-like insertion (identified by the $\hat\sigma_0$ matrix). They have been omitted in $S^{(4)}$ since those having the $\sigma_0$ insertions trivially vanish at $T=0$ when computed at zero external momenta. This is indeed a general result which follows from elementary algebra principles and holds for the whole class of diagrams having an arbitrary number of insertions of $\hat \sigma_0$ and only one insertion of $\hat \sigma_i$ (with $i=0,\dots3$). {The vanishing of these paramagnetic contributions motivated also the short-hand notation of Eq.\ \pref{jgen}-\pref{jnlgen}, where we expressed the kernel of the non-linear current only in terms of the resonant density-like response. }

{In the presence of strong disorder and/or retarded interactions the paramagnetic terms will not be exactly zero, and they could also in principle contribute to the polarization dependence of the THG signal. For example, a recent analysis of Ref.\ \cite{aoki_prb15} within the Hubbard-Holstein model has shown that the diagrams of Fig.\ \ref{Diagrams2}b become non-zero at strong coupling, with a predominance of their vertex corrections in the amplitude channel. While this could be a possible mechanism to trigger the optical visibility of the Higgs mode, this class of diagrams have a full tensorial structure, so one does {\em not} expect them to be polarization independent. Since the dynamical mean-field theory approximation used in Ref.\ \cite{aoki_prb15,shimano17} is unable to study the lattice polarization dependence, no general conclusion can be drawn on the existence of a polarization-independent Higgs contribution at strong electron-phonon coupling. }Thus a precise quantification of these processes at strong disorder/interaction remains an interesting problem for future work.

\end{document}